	\newcommand{\so}{{\bf so}}
	\newcommand{\ssl}{{\bf sl}}

	\newcommand{\tensor}{\otimes}
	\newcommand{\we}{\wedge}
	\newcommand\C{{\bf C}}
	\newcommand\R{{\bf R}}
	\newcommand\Z{{\bf Z}}
	\newcommand{\be}{\begin{equation}}
     \newcommand{\ee}{\end{equation}}
     \newcommand{\ba}{\begin{eqnarray}}
     \newcommand{\ea}{\end{eqnarray}}
     \newcommand{\ban}{\begin{eqnarray*}}
     \newcommand{\ean}{\end{eqnarray*}}
	\newcommand{\A}{{\cal A}}
	\newcommand{\D}{{\cal D}}
	\newcommand{\G}{{\cal G}}
	\renewcommand{\H}{{\cal H}}
	\newcommand{\T}{{\cal T}}
     \newcommand{\maps}{\colon}
	\newcommand{\tr}{{\rm tr}}
	
	\newcommand{\Diff}{{\rm Diff}}
	\newcommand{\End}{{\rm End}}
	\newcommand{\Fun}{{\rm Fun}}
	
	\newcommand{\M}{{\cal M}}
	\newcommand{\om}{\omega}
	\newcommand{\SO}{{\rm SO}}
	\newcommand{\SL}{{\rm SL}}
	\newcommand{\SU}{{\rm SU}}

	\documentstyle[12pt]{article}

\begin{document}

\begin{center}
{\bf Spin Networks in Nonperturbative Quantum Gravity\\}
\vspace{0.5 cm}
{\it John C.\ Baez\\}
\vspace{0.3 cm}
{\small Department of Mathematics \\
	University of California\\
        Riverside CA 92521\\}
\vspace{0.3 cm}
{\tt   baez@math.ucr.edu \\}
\vspace{0.3 cm}
{\small April 15, 1995\\ }
\vspace{0.3 cm}
{\small To appear in the proceedings of the AMS Short Course\\
on Knots and Physics, San Francisco, Jan.\ 2-3, 1995 \\ }
\vspace{0.5 cm}
\end{center}

\begin{abstract} A spin network is a generalization of a knot or
link: a graph embedded in space, with edges labelled by
representations of a Lie group, and vertices labelled by
intertwining operators.  Such objects play an important role in
3-dimensional topological quantum field theory, functional
integration on the space $\A/\G$ of connections modulo gauge
transformations, and the loop representation of quantum gravity.
Here, after an introduction to the basic ideas of
nonperturbative canonical quantum gravity, we review a rigorous
approach to functional integration on $\A/\G$ in which
$L^2(\A/\G)$ is spanned by states labelled by spin
networks.  Then we explain the `new variables' for general
relativity in 4-dimensional spacetime and describe how canonical
quantization of gravity in this formalism leads to interesting
applications of these spin network states. \end{abstract}

\section{Introduction}

Spin networks are a generalization of knots and links, and
in what follows we would like to describe some of their
recent applications to quantum gravity, and also a little
bit of their role in knot theory.  Knot
theory is a well-established branch of mathematics, full of
interesting theorems, and it is easy as a mathematician to
acquaint oneself with it by reading any of a number of texts.
Quantum gravity, on the other hand, is more difficult for the
mathematician, consisting as it does mainly of unsolved and
usually imprecisely posed questions.   Thus it seems worthwhile
to start with a thumbnail sketch of quantum gravity, focusing on
an approach known as `nonperturbative canonical quantization'.

After this heuristic introduction, we give a mathematically
rigorous account of spin networks in Section 2.   For us,
spin networks will be graphs embedded in a manifold $S$,
with edges labelled by representations of a Lie group $G$
and with vertices labelled by intertwining operators.
Spin networks define gauge-invariant functions on the space
$\A$ of connections on any $G$-bundle over $S$, and
we shall show that these functions span the gauge-invariant
subspace $L^2(\A/\G)$ of a certain Hilbert space $L^2(\A)$.
In Sections 3 and 4 we go back to the
physics and give a more detailed description of the role spin
networks play in quantum gravity.

In the latter part of the twentieth  century, much energy has
been spent in a struggle to reconcile two brilliant
accomplishments of the early part of the century:  general
relativity and quantum theory.   So far there is remarkably
little to show for all this work when it comes to verified
predictions of experimental results.  Indeed, it is quite
possible that the new predictions of a theory of quantum gravity
can only be tested at extremely small length scales, far below
those that can be probed by current experimental techniques.  The
reason for this is simple dimensional analysis: from Planck's
constant $\hbar$, Newton's gravitational constant $\kappa$, and
the speed of light $c$ one can form a quantity with units of
length, the Planck length
\[       \ell_P = (\hbar \kappa/c^3)^{1/2}, \]
in a unique way (up to dimensionless constant factors).  This
works out to be about $10^{-35}$ meters.   If we are hoping to
get experimental evidence for any theory of quantum gravity in
the forseeable future, we have to hope that somehow this simple
argument is wrong.  There are certainly many phenomena already
observed, but so far unexplained, that a theory of quantum gravity
might hope to `retrodict': phenomena from particle physics,
phenomena from cosmology, even phenomena such as the
4-dimensionality of spacetime that are taken for granted in  all
current theories of physics.   Physicists have suggested many
ideas along these lines, but  none command widespread acceptance
at this time; they are all somewhere between controversial
theories and sheer lunacy.

In short, if there were several competing well-established
theories of quantum gravity, the lack of empirical evidence to
decide between them could easily be a serious problem.  Luckily
nature has been kind to us, in that we have been unable to
formulate {\it even one} theory combining general relativity and
quantum theory in a manner that is {\it mathematically
consistent,} {\it not in obvious contradiction with experiment,}
and {\it elegant.}  The importance of mathematical consistency
and non-contradiction with experiment should be obvious, but the
role of elegance deserves some comment.  First, it is always
possible to find infinitely many consistent theories that fit a
given finite set of experimental results, just as there are
infinitely many curves through a finite set of points, and
without the freedom to reject most of them as appallingly ugly,
we would never get anywhere.  Second, there is the fact that
taken separately, general relativity and quantum theory are
strikingly esthetic as pure mathematics.   This could be a mere
coincidence, but in the absence of evidence to the contrary it is
probably best to take it as a hint, and search for a theory
combining the most beautiful aspects of both in an integral
manner.  Indeed, though the `unreasonable effectiveness'  of
elegant mathematics is still a great mystery, it is also an
empirical fact, so even as hard-nosed empiricists we should heed
it.

Many physicists, particularly in string theory, have argued that
it is impossible to reconcile general relativity and quantum
theory without taking all the other forces and particles into
account. Here however we concentrate on a diametrically opposite
approach, which tries to construct a quantum version of the {\it
vacuum} Einstein equations --- the theory of gravity with {\it
no} matter around.  We shall not attempt to discuss the relative
merits of the two approaches, but simply note the curious fact
that some of the same mathematics shows up in both, possibly for
some deep reasons that would be good to understand \cite{Baez4}.

In what follows we need to assume a nodding acquaintance with
some differential geometry.   Let us take as our spacetime a
4-manifold $M$ diffeomorphic to $\R \times S$, where $\R$
represents time and the 3-manifold $S$ (compact and orientable to
simplify the discussion) represents space.  The vacuum Einstein
equation concerns a Lorentzian metric $g$ on $M$. Given such a
metric there is a unique metric-preserving, torsion-free
connection $\Gamma$, the Levi-Civita connection. The curvature of
this connection is described by the Riemann tensor $R$, which in
local coordinates is written  ${R^{\alpha}}_{\beta \gamma
\delta}$.   A certain amount of information about the curvature
of spacetime is thus contained in the Ricci tensor $R_{\mu \nu} =
{R^{\alpha}}_{\mu \alpha \nu}$, and the vacuum Einstein
equation says simply that
\[              R_{\mu \nu} = 0.\]

To understand why precisely {\it these} are the equations
describing the curvature of empty spacetime, though, it is
necessary to step back and consider the case where there is
matter present.  In the presence of matter, the density of energy
and momentum and their rate of flow in different spatial
directions are summarized by a symmetric tensor $T_{\mu \nu}$.
The sense in which energy and momentum are conserved in general
relativity is quite subtle, but the simplest way of stating this
conservation principle is that $T_{\mu\nu}$ is `divergence-free':
\[        \nabla^\mu T_{\mu \nu} = 0 ,\]
where $\nabla$ is the operator of covariant differentiation
corresponding to the Levi-Civita connection.  If we wish to say
that spacetime is curved by energy and momentum, it is natural
to try some equation like
\[             C_{\mu \nu} = T_{\mu\nu} \]
where $C$ is built up from the Riemann tensor in some simple way.  However,
are constrained by the fact that $C$ must be symmetric and
divergence-free.  The simplest thing to try is some multiple of
\[            G_{\mu \nu} = R_{\mu \nu} - {1\over
2}R^\alpha_\alpha g_{\mu \nu} ,\]
because this is symmetric and, by the Bianchi identity, also divergence-free.
In fact, Einstein's equation in the presence of matter says that
\[            G_{\mu \nu} = 8 \pi \kappa T_{\mu \nu} .\]
But in a vacuum $T_{\mu \nu} = 0$, so we obtain
\[           G_{\mu \nu} = 0 ,\]
which turns out to be equivalent to the vanishing of the Ricci
tensor.

To see how the vacuum Einstein equation describes
the {\it dynamics} of the metric, it
is useful to split spacetime into space and time.  The
manifold $M$ is diffeomorphic to $\R \times S$, but not in any
canonical way.  Nonetheless, let us go ahead and pick a
diffeomorphism between them.  This lets us transfer the usual
coordinate function on $\R$ to a `time' function $x^0$
on $M$, so we can talk about the metric on space at time
$t$, that is, the metric $g$ restricted to the slice
\[              S_t = \{x^0 = t\} \subseteq M .\]
Assume we can choose the slices $S_t$ to be spacelike, meaning
that the restriction of $g$ to each one is a Riemannian metric.
We can only do this if $g$ is `globally hyperbolic', but
physically this is a reasonable condition.   Then, near
any point $p \in M$, we can find local coordinates $x^\mu$ such
that $x^0$ is the above time function, the vector field
$\partial_0$ is normal to the slices,  and the vector fields
$\partial_i$ corresponding to the remaining `space' coordinates
$x^i$ ($i = 1,2,3$) are tangent to the slices.  One does not
really need to use coordinates with these special properties, but
it simplifies the discussion below.

Since $G_{\mu\nu}$ is a symmetric $4 \times 4$ matrix, the vacuum
Einstein equation really has 10 components.  Splitting spacetime
into  space and time lets us interpret these components as
follows. The equation $G_{00} = 0$ and the 3 equations $G_{0i} =
0$ are  {\it constraints} on the metric on space, $g_{ij}$,  and
its first time derivative $\dot g_{ij}$.  In other words,
$G_{00}$ and $G_{0i}$ can be computed on the slice $S_t$ in terms
of the `initial data' $g_{ij}$ and $\dot g_{ij}$ on that slice,
and for there to be a solution of Einstein's equation having given
initial data, the initial data must satisfy the constraints
$G_{00} =  G_{0i} = 0$.  The remaining 6 equations $G_{ij} = 0$
are {\it evolutionary equations} involving the second time
derivative $\ddot g_{ij}$.  They allow us to compute how the
metric evolves in time, given that the constraints hold
initially.

In fact, the constraints have a simple meaning, which turns out
to be very important.  To understand this meaning, however, we
need a brief detour into classical mechanics.  It is worth
recalling the simplest classical mechanics problem of all, the
motion of a particle on the line in a potential $V$. The
particle's position is a point $q$ in the space $\R$, and one calls this
space of possible positions the `configuration space'.   The
motion of the particle is determined by its position $q$ and
velocity $\dot q$ at time zero, but it is often handier to
work not with its velocity but with its momentum $p = m\dot q$,
where $m$ is its mass.  The position and momentum determine a
point $(q,p)$ in the `phase space' $T^\ast \R$, since the
momentum is most naturally regarded as a cotangent vector to the
configuration space.   This phase space has a closed
nondegenerate 2-form, or `symplectic structure', on it, given by
\[  \om = dp \wedge dq .\]
The energy, or Hamiltonian, of the particle is a function on
phase space, given by the sum of kinetic and potential energy:
\[   H = {p^2 \over 2m} + V(q) .\]
The Hamiltonian gives rise to a vector field $v_H$ on phase space by
\[      \om(\cdot,v_H) = dH .\]
This vector field generates a one-parameter group of
diffeomorphisms of phase space, or `flow', which
describes the time evolution of the particle.  For short, one
says that the Hamiltonian generates time evolution.
Indeed, quite often in classical mechanics the phase space is a
symplectic manifold and time evolution is generated by the
Hamiltonian in this manner.  And quite often the phase space is the
cotangent bundle of some configuration space; cotangent bundles
are equipped with a canonical symplectic structure.

In general relativity, the  quantity analogous to the `position'
is the metric $g_{ij}$ on  space, and we shall henceforth write
this as $q_{ij}$ to emphasize the analogy.  The configuration
space of general relativity is thus the space $\M$ of all
Riemannian metrics on $S$.  The quantity analogous to the
`velocity' is the time derivative $\dot q_{ij}$.   Following
certain standard recipes, the analog of the momentum works out to
be
\[  p^{ij} = {1\over 2} (\det q_{ij})^{1/2} (\dot q^{ij} - \dot q_k^k
q^{ij}) . \]
The phase space of general relativity is the cotangent bundle
$T^\ast \M$, and the pair $(q_{ij},p^{ij})$
determines a point in this phase space.

Since one can write the constraints $G_{00}$ and $G_{0i}$ in
terms of $q_{ij}$ and $p^{ij}$, one can think of these
constraints as functions on $T^\ast \M$.   From this point of
view, they play a dual role \cite{FM}.  First, as already noted,
for the pair $(q_{ij},p^{ij})$ to determine a  solution of
Einstein's equation, it must lie on the subspace of $T^\ast \M$
where the constraints vanish.  Second, and more
subtly, the constraints generate physically important flows on
phase space.

For example, if we integrate $G_{00}$
over $S$, we obtain a function on $T^\ast \M$ which generates
time evolution with respect to the time coordinate $x^0$.
For this reason physicists call this constraint the `Hamiltonian
constraint', and use the notation
\[     \H = G_{00}  \]
in this context.   On the other hand, if we take a vector field
$N$ on $S$ and integrate $N^i G_{0i}$ over $S$, we obtain a
function on $T^\ast \M$ generating a flow which is the same as
that induced by the one-parameter group of diffeomorphisms of $S$
generated by $N$.  Physicists call this constraint the
`diffeomorphism constraint', and write it as
\[       \H_i = G_{0i} .\]
It is important to note, however, that
both $\H$ and $\H_i$ correspond to one-parameter groups
of diffeomorphisms of spacetime: $\H$ corresponds to
diffeomorphisms that `push $S$ forwards in time', while $\H_i$
corresponds to diffeomorphisms that map $S$ to itself.

Now let us turn the problem of quantizing general relativity,
that is, guessing  a quantum theory that reduces to Einstein's
equation in the limit $\hbar \to 0$.  Again, it is good to recall
the  example of a particle on the line.  In the quantum theory of
a particle on a line, the particle's state is given by a
`wavefunction', that is,  an $L^2$ function $\psi$ on the
configuration space $\R$. If we assume $\psi$ is normalized so
that $\|\psi\|= 1$,  the probability of finding the particle in
any open subset $U \subseteq \R$ is then given by
\[        \int_U |\psi(x)|^2 \, dx  .\]
Now, classically the Hamiltonian is
\[     H(p,q) = {p^2\over 2m} + V(q) ,   \]
and to quantize the particle one simply replaces
$p$ and $q$ in this formula with certain operators $\hat p, \hat q$
on $L^2(\R)$:
\[     (\hat p \psi)(x) = {\hbar \over i}{d\over dx} \psi(x),\qquad
    (\hat q \psi)(x) = x \psi (x),\]
obtaining an operator $\hat H$, the quantum Hamiltonian:
\[     (\hat H \psi)(x) = -{\hbar^2\over 2m} {d^2 \over dx^2}
\psi(x) + V(x) \psi (x)  .\]
The Hamiltonian $\hat H$ then describes the evolution of the
wavefunction in time via Schr\"odinger's equation:
\[          i\hbar {d\over dt} \psi = \hat H \psi .\]

This recipe for replacing $p$ and $q$ by operators $\hat p$ and
$\hat q$ is known as `canonical quantization'. If we try to copy
this recipe in the case of general relativity, we quickly notice
some serious problems.  In the case of the particle on the line,
the `configuration space' in which $q$ lies is just the real
line, and we can define $L^2(\R)$ using Lebesgue measure.  In the
case of general relativity, the configuration space is the space
$\M$ of all Riemannian metrics on $S$.  This is an {\it
infinite-dimensional} manifold, and there is no good notion of
`Lebesgue measure' on such spaces, so defining the Hilbert space
$L^2(\M)$ is not so easy.

Of course, this problem will arise whenever we try to canonically
quantize a theory with infinite-dimensional configuration space.
Thus it occurs, not only in quantum gravity, but in other quantum
field theories.   When quantizing linear equations, where the
configuration space is an infinite-dimensional vector space $\cal
V$, one can use an infinite-dimensional analog of a Gaussian
measure to define $L^2(\cal V)$, and the resulting theories make
fine mathematical sense.   This strategy is not sufficient by
itself to deal with nonlinear equations such as Einstein's
equation, however.  Indeed, $\M$ is not even a vector space.

Another problem is that unlike the particle on the line,  whose
position and velocity at a given time are arbitrary, general
relativity involves {\it constraints}.   One should take these
constraints into account when quantizing the theory, but the
correct way to do so is not at all clear!  Indeed, much ink has
been spilled concerning the quantization of constrained systems,
and we cannot go into all the nuances of this issue here.
Instead, let us simply give an oversimplified account of Dirac's
approach to constraints  \cite{Dirac}, as applied to gravity by
DeWitt \cite{DeWitt} in the late 1960s.

Suppose we could somehow get ahold of a Hilbert space $L^2(\M)$.
The idea is that not all vectors in this so-called `kinematical' state space
represent physical states of quantum gravity, but
only those satisfying certain quantum versions of the
constraints.   To describe these quantized
constraints, first we try to define operators $\hat q_{ij}$
and $\hat p^{ij}$ on $L^2(\M)$ by the formulas
\[     (\hat p^{ij} \psi)(q) = {\hbar \over i}{\partial \over
\partial q_{ij}} \psi(q),\qquad
    (\hat q_{ij} \psi)(q) = q_{ij} \psi (q),\]
where $q \in \M$.  These formulas are merely heuristic, and part
of the problem is making good enough sense of them for the task
at hand.    For example, in analogy with $p^{ij}$ and $q_{ij}$ in
the first place, $\hat p^{ij}$ and $\hat q_{ij}$ are really
something like operator-valued functions on the subset of the
spacelike slice contained in our coordinate chart.  When one
tries to be rigorous one discovers that they are
operator-valued distributions.  As noted above, one can work out formulas
for the constraints $\H$ and $\H_i$ in terms of $p^{ij}$ and
$q_{ij}$.  The idea is then to take these formulas, and replace
all appearances of $p^{ij}$ and $q_{ij}$ in them by $\hat p^{ij}$
and $\hat q_{ij}$, obtaining operator-valued distributions $\hat \H$
and $\hat \H_i$.   For a wavefunction $\psi \in L^2(\M)$ to
represent a physical state of quantum gravity, the following
quantum versions of the constraint equations should hold:
\[         \hat \H \psi = 0, \qquad \hat \H_i \psi = 0.\]

There are some deep conceptual problems associated with
this approach to quantum gravity.   For example, instead of
Schr\"odinger's equation, in which the Hamiltonian describes
the time evolution of the wavefunction, all we have is the so-called
`Wheeler-DeWitt equation' $\hat \H \psi = 0$.
What does this mean?  Where has the {\it dynamics} of quantum
gravity gone, if physical states are annihilated by the operator
whose classical counterpart generates time evolution?  Briefly,
the answer appears to be this:  the constraint equations say that
$\psi$ only describes {\it diffeomorphism-invariant} information
about the world.  In other words, we cannot use $\psi$ to answer
questions about what is happening at a certain point whose
location is specified using a coordinate system, such as `what is
the metric at the point $x^\mu =  a^\mu$?'.   Unfortunately, we
are not used to doing physics without asking such questions!
This is known as the `problem of time' in quantum gravity
\cite{Isham}.

Just as bad as the conceptual problems are the sheer technical
problems involved in making mathematical sense out of DeWitt's
strategy for quantizing gravity.  In fact, nobody has yet
succeeded.   It is worth noting that there are other constrained
nonlinear equations with infinite-dimensional configuration
spaces where many of the same problems show up: for example, the
Yang-Mills equation, various versions of which appear to describe
all the forces {\it except} gravity.  Quantizing these in a
rigorous way is {\it also} extremely difficult, but physicists have
had much more success with them at the practical level.  Let us
briefly summarize the difference between the two cases.

Physicists often quantize nonlinear field theories by treating
the nonlinearities as `small perturbations' of some linear
equation.  There are also perturbative methods for dealing with
constraints.  These methods can be problematic, mathematically
speaking, but unless afflicted by an unusual desire
for rigor, physicists are often happy to do perturbation theory
using formal power series in some coupling constant that measures
the nonlinearity.  Due to the nonrigorous nature of these
computations,  the coefficients of these power series usually
come out to be ill-defined unless one performs some clever
maneuvers known as renormalization.  One says the theory is
renormalizable if these maneuvers work, but renormalizability by
itself does not mean that the power series are convergent, or
even asymptotic.  Thus, strictly speaking, renormalizability is
not a sufficient condition for a nonlinear quantum field theory
to make rigorous sense.   Nor, on the other hand, is
renormalizability a {\it necessary} condition to be able to
quantize a nonlinear wave equation, for there are also other
`nonperturbative' approaches.   In practice, however, particle
physicists often restrict themselves to  renormalizable theories,
make physical predictions using the first few terms in the power
series, and compare these predictions with experiment to see if
the theory is on the right track.

For the Yang-Mills equation, and indeed for the whole `Standard
Model' of particles and forces other than gravity, this strategy
has been quite successful.  But the strategy fails miserably with
quantum gravity, for the theory resists all attempts at
renormalization.   As it turns out, this is closely related to
the fact that the constraints are not polynomials in the basic
`position' and `momentum' variables, $q_{ij}$ and $p^{ij}$, and
their derivatives.   Instead, they contain nasty factors of
$(\det q_{ij})^{-1/2}$, essentially because the formula for
$p^{ij}$ contains a factor of $(\det q_{ij})^{1/2}$.   Standard
methods for replacing $q_{ij}$ with the operator $\hat q_{ij}$
lead to all sorts of problems when applied to such non-polynomial
expressions.  (Similar problems arise in the path-integral
approach which is more commonly used in renormalization.)  To
overcome the nonrenormalizability of quantum gravity, particle
physicists constructed ever more complicated models containing
gravity and other forces, in order to cancel out unruly
infinities.  This led to the study of supergravity and
eventually superstring theory, which attempts to model {\it all}
known forces and particles, and unfortunately many unknown ones
as well.

Meanwhile, many general relativists were suspicious of the whole
idea of perturbatively quantizing gravity.   This amounts to
treating all metrics as small perturbations of a fixed
`background metric', usually taken to be the flat Minkowski
metric on $\R^4$.  Dimensional analysis, however, suggests that
one should expect more and more extreme quantum fluctuations of
the metric at  smaller and smaller length scales, becoming very
significant at about the Planck length.  So perhaps one should
really adopt some {\it nonperturbative} approach to canonical
quantum gravity.

Saying this is easy: the problem is actually doing it.
In principle, the most direct way would be to make DeWitt's
approach into rigorous mathematical physics without reference to
a fixed background metric.   But attempting to do this led
immediately into a quagmire which for two decades seemed
impassable.  It was not until the late 1980s, when Ashtekar
\cite{Ash1} found a clever change of variables, that a way around
began to seem possible.

We postpone a detailed discussion of these `new variables'  to
Section 3.  For now, let us simply note that in terms of them,
the configuration space of general relativity space is not the
space of {\it metrics} on $S$, but the space of {\it connections}
on some $\SL(2,\C)$ bundle over $S$.  Since $\SL(2,\C)$ is  a
complex Lie group, this configuration space is a complex
manifold. It turns out that in the  quantum theory, kinematical
states should be holomorphic functions on this configuration
space.   However,  using the fact that $\SL(2,\C)$ has $\SU(2)$ as
a real form, one expects the kinematical state space to be
isomorphic to $L^2(\A)$, where $\A$ is the space of connections
on a certain $\SU(2)$ bundle over $S$.   Thus, at least naively,
one expects the physical states of quantum gravity in the new
variables formalism to be wavefunctions $\psi \in L^2(\A)$
annihilated by certain constraints.   These constraints include
Hamiltonian and diffeomorphism constraints as before, which express the
invariance of $\psi$ under diffeomorphisms of spacetime, but also
a `Gauss law' constraint which expresses the invariance of $\psi$
under gauge transformations.

Now the configuration space of the Yang-Mills equation is also a
space of connections, and the only constraint in Yang-Mills
theory is the Gauss law.  Thus in terms of the new variables,
canonical quantum gravity is very similar to $\SU(2)$
Yang-Mills theory with some extra constraints!  This is one of
the main advantages of the new variables: they allow techniques
from Yang-Mills theory to be imported to quantum gravity.
Another more technical advantage is that in the new variables,
the constraints  are polynomial functions of the analogs of
position and momentum (and their derivatives).    Because the
nonrenormalizability of quantum gravity in the traditional metric
formulation was closely related to the non-polynomial nature of
the constraints, this created a lot of excitement.  Unfortunately,
at the present time the Hamiltonian constraint still presents
thorny problems.

Shortly after the discovery of the new variables, Rovelli and
Smolin \cite{RS} used them to develop a `loop representation' of
quantum gravity, and this is when the relationship to knot theory
became very apparent.   The idea of the loop representation of a
gauge theory had been developed by Gambini and Trias \cite{GT},
and basically it consists of writing all the equations one can in
terms of `Wilson loops'.  Wilson loops are certain  functions on
the space of connections: at a point $A \in \A$, the value of the
Wilson loop is just the trace of the holonomy of $A$ around some loop
$\alpha$ in $S$, taken in some (finite-dimensional) representation
$\rho$ of $G$, written:
\[    \tr(\rho(T \exp{\int_\alpha A}))  .\]
One reason why physicists like them is that they are
invariant under gauge transformations, and one expects the
physically observable aspects of a gauge theory to be gauge-invariant.

Rovelli and Smolin argued that a state $\psi$ of
quantum gravity should give rise to link invariant $\hat \psi$,
the `loop transform' of $\psi$,
whose value on the link with components given by the loops
$\alpha_1, \dots, \alpha_n$ is
\[     \hat \psi (\alpha_1, \dots, \alpha_n) =
\int_{\A} \,\prod_{i=1}^n \tr(\rho((T \exp{\int_{\alpha_i} A}))\,\psi(A)\,
\D A, \]
where $\rho$ is the fundamental representation
and ${\cal D}A$ is the purely formal `Lebesgue measure' on $\A$.
The reason $\hat \psi$ should be
a link invariant is that the diffeomorphism constraint says
$\psi$ is invariant under diffeomorphisms of $S$!   Of course,
this argument is merely heuristic, owing to the mysterious nature
of $\D A$, but it is no worse than much of the reasoning
in quantum gravity.   Indeed, a similar sort of argument
led Witten \cite{Witten} to discover the relation between
knot theory and another quantum field theory, Chern-Simons theory.
In fact, the link
invariant coming from $\SU(2)$ Chern-Simons theory, namely
the Kauffman bracket, appears to be the loop
transform of a state of quantum gravity `with cosmological
constant', meaning that the Einstein equation has been modified
to give the vacuum a nonzero stress-energy tensor.   We will not
go into this here, since there are a number of expository
treatments already \cite{Baez4,Baez3,BM,Br,Pullin}, but
certainly it is one of the main reasons for interest in the
interface between knots and quantum gravity.

There is much here that needs to be made more precise in the
following sections.  However, we are at least in a position now
to describe what needs to be done.  First, we need to develop
integration theory on $\A$, in order to escape the use of purely
formal entities like the `Lebesgue measure' $\D A$. In Section 2
we do this when $\A$ is the space of smooth connections on  any
principal $G$-bundle $P$ over a manifold $S$, where $G$ is a
compact connected Lie group and $S$ is real-analytic.   There is
an especially nice `generalized measure' on $\A$ which is a
substitute for the nonexistent Lebesgue measure.  Using it we can
define $L^2(\A)$, and it turns out that the gauge-invariant
subspace of $L^2(\A)$ is spanned by `spin network states'.  These
are described by graphs analytically embedded in $S$, with
oriented edges labelled by  representations of $G$, and with
vertices labelled by intertwining operators from the tensor
product of representations labelling `incoming' edges to the
tensor product of representations labelling `outgoing' edges.
Equipped with this mathematical technology, we then return to
quantum gravity.

\section{Spin Networks}

In
quantum field theory computations, physicists often try to do
integrals over a space $\A$ of connections using a
strange thing they call `Lebesgue measure' on $\A$,  usually
written $\D A$.   This exploits the fact that $\A$ is an affine
space, or, arbitrarily choosing one point as the origin, a vector
space.  The idea is that one should be able to pick a basis for
$\A$, thus setting up an isomorphism
\[ \A \cong \R^\infty ,\]
and then, working in the coordinates defined by this basis, let
\[ \D A = \prod_{i=1}^\infty dx_i.  \]
There are many problems with this idea, however!  First, while
$\A$ can indeed be identified with an infinite-dimensional
vector space, a basis of $\A$ does not really give an isomorphism
between $\A$ and an infinite product of copies of $\R$.  Second,
there is no good theory of infinite products of arbitrary measure
spaces.  Of course one can be more sophisticated about these
issues, but the fact remains that there is no such thing as
`Lebesgue measure' on an infinite-dimensional vector space. Thus
it is not surprising that when physicists actually do integrals
over $\A$ using `Lebesgue measure' they often get infinite or
ill-defined answers until they perform various sneaky tricks.

On the other hand, there is a different way of doing these
integrals which is used in lattice gauge theory.  Lattice gauge
theory is an approximation to gauge theory on $\R^n$ in which one
replaces the continuum by an infinite graph having the points of
a lattice as vertices and the line segments between
neighboring vertices as edges.  A `connection' on the lattice
is simply an assigment of an element $g_e$ of the gauge group $G$ to each edge
$e$ of the graph, representing the effect of parallel transport
along $e$.   The space of connections in this context is thus
very different: it is
\[            \A \cong G^\infty ,\]
not an infinite product of copies of $\R$, but an infinite
product of copies of $G$!  Now, while an infinite product of
arbitrary measure spaces is ill-defined, an infinite product of
{\it probability} measure spaces {\it is} well-defined.
(Recall that a probability measure space is a space $X$ with
positive measure $\mu$ such that $\int_X \mu = 1$.)
If $G$ is compact, it is equipped with a very natural probability
measure, namely Haar measure, the unique left- and
right-invariant Borel measure $dg$ with $\int_G dg = 1$.
So in lattice gauge theory with compact gauge group,
one can work with the measure
\[         \D A = \prod_e dg_e \]
on $\A$, instead of the nonexistent `Lebesgue measure'.  This is precisely
what is done.

Now, while lattice gauge theory is easier to make rigorous, it
has the disadvantage of being only an approximation to the
continuum theories we are really interested in.   Indeed, in
practice much work in lattice gauge theory is computational in
nature.  Here one uses not a lattice but a finite graph, so that
$\A$ becomes a {\it finite} product of copies of $G$, and
one numerically calculates integrals over $\A$ by Monte Carlo
methods.   To apply these results to gauge theory on $\R^n$ one
must then investigate not only the `continuum limit' in which the lattice
spacing goes to zero, but also the `large-volume limit'.

Approximating the continuum by a lattice or a fixed finite graph
is particularly distressing in the case of general relativity,
which is so intimately connected to the differential geometry of
manifolds.  It would be nice if one could have ones cake and eat
it too, working with connections in the continuum context and
preserving diffeomorphism-invariance, but still thinking of the
space of connections as something like a product of copies of
$G$.  This is precisely what the theory of `generalized measures'
on the space of connections seeks to achieve.

The traditional approach to measure theory  is to pick a
$\sigma$-algebra of `measurable' subsets of some space $X$,
assign measures to them in a manner satisfying some axioms, and
then define a vector space of `integrable'
complex-valued functions on $X$.   Then one figures out how to
integrate functions in this space, shows that integration is
linear, and shows how to pass limits through integrals under
certain conditions.  Then, typically, one forgets the proofs of
these results and simply uses them.   A more modern approach is
to shortcut this process and simply choose a space $\Fun(X)$ of
functions on $X$, equip it with a topology, and define a
`generalized measure' $\mu$ on $X$ to be a continuous linear
functional
\[                \mu \maps C \to \C  ,\]
writing $\mu(f)$ as $\int_X f\, d\mu$ solely out of deference to
tradition.   Actually, of course, the modern approach complements
the old approach rather than replaces it; they are related by
many useful theorems, some of which we have summarized elsewhere
\cite{Baez}.

Now suppose that $P$ is a principal $G$-bundle over $S$, where
$G$ is compact and connected, and let $\A$ be the space of smooth
connections on $P$.  To define generalized measures on $\A$ we
need to choose the space $\Fun(\A)$ of functions we want to
integrate.  Following the idea of the loop representation, for
example, we could choose $\Fun(\A)$ to be the algebra of
functions on $\A$ generated by all Wilson loops, or  a completion
of this algebra in some topology.   In fact, this is how Ashtekar
and Isham \cite{AI} proceeded in their original attempt to use
ideas from the loop representation to set up a rigorous
integration theory on $\A$.  Subsequent work by Ashtekar,
Lewandowski, and the author \cite{AL,AL2,Baez,Baez2,L} improved
things in a number of ways.  For one, it turns out to be simpler
to let $\Fun(\A)$ contain arbitrary continuous functions of the
holonomy of $A$ along {\it paths} in $S$.   Also, working with
smooth paths turns out to be a nuisance, because they can
intersect each other in horribly complicated ways.   For this
reason we shall follow Ashtekar and Lewandowski and require $S$
to be a real-analytic manifold, and work with piecewise
real-analytic paths.  In a sense this is not so drastic, because
every compact smooth manifold can be given a real-analytic
structure.   One pays a price, however, since one is no longer
doing differential geometry in the category of smooth manifolds.
Eventually it would be nice to understand the smooth case too.

In any event, let us define $\Fun(A)$ as follows.  If $\gamma
\maps [0,1] \to S$ is
a piecewise analytic path, let $\A_\gamma$ denote the
space of smooth maps $F\maps P_{\gamma(0)} \to P_{\gamma(1)}$ that
are compatible with the right action of $G$ on $P$:
\[            F(xg) = F(x)g .\]
Note that for any connection $A \in \A$, the parallel transport
map
\[        T \exp {\int_\gamma A}: P_{\gamma(0)} \to P_{\gamma(1)} \]
lies in $\A_\gamma$.  If we fix a
trivialization of $P$ at the endpoints of $\gamma$,
we can identify $\A_\gamma$ with the group $G$.   This makes
$\A_\gamma$ into a compact manifold in a manner which one can
check is independent of the trivialization.   Let
$\Fun_0(\A)$ be the algebra of functions on $\A$ generated by
those of the form
\begin{equation}
 \psi(A) = f(T \exp{\int_{\gamma_1} A},\dots,T \exp{\int_{\gamma_n} A})
\label{form} \end{equation}
where $\gamma_1, \dots, \gamma_n$ are piecewise analytic paths
in $S$, and $f$ is a continuous function on $\A_{\gamma_1} \times
\cdots \times \A_{\gamma_n}$.  Then let $\Fun(\A)$ be
the completion of $\Fun_0(\A)$ in the sup norm:
\[          \|\psi \|_\infty = \sup_{A \in \A} |\psi(A)|. \]
It is easy to check that the Wilson loops lie in this algebra; in
fact, they lie in $\Fun_0(\A)$.

A generalized measure on $\A$ is then defined to be a continuous
linear functional on $\Fun(\A)$.  This may seem rather abstract,
so let us see how to get our hands on one. There exists a nice
general recipe for constructing {\it any} generalized measure on
$\A$, which we have described elsewhere \cite{Baez2,Baez3}, but
for now let us concentrate on a simple and important example: the
uniform generalized measure $\mu_u$.  This is a kind of
replacement for the nonexistent `Lebesgue measure' on $\A$, and
as we shall see, it is closely modelled after the measure $\D A$
in lattice gauge theory.

To define $\mu_u$, first we
define it as a linear functional on $\Fun_0(\A)$.
Because we are working with real-analytic paths, it
turns out that any $\psi
\in \Fun_0(\A)$ can be written in the form given by equation (\ref{form})
with the paths  $\{\gamma_i\}$ forming an {\it embedded graph} in
$S$.   By this we mean that each path $\gamma_i \maps [0,1] \to
S$ is one-to-one and restricts to an embedding of $(0,1)$, and the
images $\gamma_i[0,1]$ and $\gamma_j[0,1]$ intersect, if at all,
only at their endpoints when $i \ne j$.
For functions $\psi$ written in this special form we define
\[ \int_\A \psi\, d\mu_u =
\int_{G^n} f(g_1, \dots, g_n) dg_1 \cdots dg_n .\]
Here we are using trivializations of $P$ at the endpoints of
the paths $\gamma_i$ to identify $\A_{\gamma_1} \times \cdots
\times \A_{\gamma_n}$ with $G^n$, but the right-hand side is
independent of the choice of trivializations, because Haar
measure is right- and left-invariant.  All the real work goes
into checking that the right-hand side does not depend on {\it how} we
wrote $\psi$ in this special form involving an embedded graph.
Given that, it is easy to check that
$\mu_u$ is a linear functional on $\Fun_0(\A)$.
The bound
\[       |\int_\A \psi \, d\mu_u|\le \|\psi\|_\infty \]
then holds because Haar measure is a probability
measure.   Since $\Fun_0(\A)$ is dense in $\Fun(\A)$, this bound
implies that $\mu_u$ extends uniquely to a continuous linear
functional on all of $\Fun(\A)$, which we again call $\mu_u$.
This is the uniform generalized measure on the space of connections!

If we examine what we have just done, the relationship to lattice
gauge theory should become clear.  Instead of working with a
fixed graph embedded in $S$ as one does in lattice gauge theory,
we have considered {\it all possible} embedded graphs $\gamma =
\{ \gamma_i\}$ in $S$.  Each one of these indeed has the topology
of a graph with the paths $\gamma_i$ as edges and the endpoints
$\gamma_i(0), \gamma_i(1)$ as vertices.  Following the spirit of
lattice gauge theory, we can define a finite-dimensional space of
`connections' on $\gamma$,
\[   \A_\gamma = \A_{\gamma_1} \times \cdots \times \A_{\gamma_n},
\]
on which there is a natural measure, namely a product of copies
of Haar measure, one copy for each edge of $\gamma$.  To do the
integral of a function that only depends on the holonomies along
the edges of $\gamma$, we simply use this natural measure.  The
key fact is that although a function  $\psi \in \Fun_0(\A)$ can
be expressed in terms of many different embedded graphs in this
way --- after all, if it can be expressed in terms of a graph
$\gamma$, it can also be expressed in terms of any graph
$\gamma'$ containing $\gamma$ --- the answer we get for the
integral of $\psi$ is independent of this choice.   This is why
we obtain a well-defined linear functional $\mu_u$ on $\Fun_0(\A)$, which
then extends to all of $\Fun(\A)$.     In short, we are getting a
generalized measure on $\A$ as a kind of `limit' of measures on
the spaces of connections on all graphs embedded in $S$!  This can be
made perfectly precise using the language of projective limits
\cite{AL2}.

We can now define the Hilbert space $L^2(\A)$ as the completion
of $\Fun(\A)$ with respect to the norm
\[         \|\psi\|_2 = (\int_\A |\psi|^2 \, d\mu_u)^{1/2}  .\]
In the special case where $G = \SU(2)$ and $S$ is a compact
oriented 3-manifold, this Hilbert space will be our space of
`kinematical states' for quantum gravity in the new variables
formalism.   The Gauss law constraint then amounts to requiring
that the state $\psi \in L^2(\A)$ be invariant under gauge
transformations, so it is important to find gauge-invariant
vectors in $L^2(\A)$.   Finding these vectors is actually of
interest no matter what $G$ and $S$ happen to be.   It turns out
that these vectors can also be  thought of as wavefunctions on
the quotient of $\A$ by the group $\G$ of smooth gauge
transformations \cite{Baez3}, so we will denote the space of
gauge-invariant vectors in $L^2(\A)$ by $L^2(\A/\G)$.

The most obvious examples of vectors in $L^2(\A/\G)$
are the Wilson loops.  If we have an analytic loop $\alpha$ in
$S$ and a representation $\rho$ of $G$, the function
\[    \psi(A) = \tr(\rho(T \exp {\int_\alpha A}))  \]
is gauge-invariant, and it lies in $L^2(\A)$.
More generally, any product of Wilson loops is a gauge-invariant
element of $L^2(\A)$.

More generally still, we can get vectors in $L^2(\A/\G)$ from
`spin networks' \cite{Baez5}.   Take a graph $\gamma$ embedded in $S$ and
label each of its edges $e$ with a representation $\rho_e$ of
$G$.  Let
\[      H_e(A) = \rho_e(T \exp \int_e A) .\]
If we trivialize $P$ at the vertices of $\gamma$, and pick a
basis for $\rho_e$, we can think of $H_e(A)$ as a matrix
$H_e(A)^i_j$.  Now form the tensor product of all these matrices,
one for each edge of $\gamma$.  We get a big tensor $H(A)$ having
one superscript and one subscript for each edge; it is too ugly
to bother writing down, but we hope the reader gets the idea.

Next, for each vertex $v$ of $\gamma$
let $S(v)$ be the set of edges having $v$ as `source' --- where
the source of an edge $\gamma_i$ is defined to be $\gamma_i(0)$ ---
and let $T(v)$ be the set of edges having $v$ as `target' ---
where the target of $\gamma_i$ is $\gamma_i(1)$.    For
each vertex $v$, pick an intertwining operator
\[  I_v \maps \bigotimes_{e \in T(v)} \rho_e
\to \bigotimes_{e \in S(v)}\rho_e .\]
We can think of $I_v$ as a tensor with one superscript for each
edge $e \in T(v)$ and one subscript for each edge $e \in S(v)$.
Then form the tensor product of all these tensors $I_v$, one for
each vertex.  We get a big tensor $I$.  Then form the tensor
product of $H(A) \tensor I$.  Note that each superscript of $H(A)$
corresponds to a particular subscript of $I$ and vice versa,
because each edge of $\gamma$ lies in $S(v)$ for one vertex $v$
and lies in $T(w)$ for one vertex $w$.  So we can contract the
tensor $H(A) \tensor I$ to get a number, which of course depends
on $A$.  This is our `spin network state' $\psi(A)$.   Note that a
Wilson loop is just a special case of a spin network with only
one edge and one vertex, with the intertwining operator taken to
be the identity operator.

One can show directly from this explicit definition that the spin
network states are gauge-invariant and lie in $L^2(\A)$, in fact
in $\Fun_0(\A)$.  But we want to show more: we want to show they
span the whole space of gauge-invariant vectors in $L^2(\A)$.
For this a more abstract approach is better.

The group $\G$ acts on $\A$, and in fact there is a unitary
representation of $\G$ on $L^2(\A)$ given by
\[          g\psi(A) = \psi(g^{-1}A)  .\]
To see this it is useful to introduce work `one embedded graph at a
time'.  One can show that for any embedded graph $\gamma$,
$L^2(\A_\gamma)$ can be identified with the smallest
closed subspace of $L^2(\A)$ containing all functions of the form
given in equation (\ref{form}).
If we do a gauge transformation $g \in \G$ on the connection
$A$, the holonomy along any edge $\gamma_i$ transforms to
\begin{equation}  T \exp\int_{\gamma_i} gA = g(\gamma_i(1)) \,
(T \exp\int_{\gamma_i} A) \, g(\gamma_i(0))^{-1},  \label{tran} \end{equation}
so if $\psi$ lies in $L^2(\A_\gamma)$, so does $g\psi$.
Thus we get a representation of $\G$ on each subspace
$L^2(\A_\gamma)$, and these representations are unitary
because Haar measure is left- and right-invariant.  Since the union of
the subspaces $L^2(\A_\gamma)$ is dense in $L^2(\A)$, we obtain
a unitary representation of $\G$ on all of $L^2(\A)$.

Let $L^2(\A_\gamma/\G_\gamma)$ denote the gauge-invariant
subspace of $L^2(\A_\gamma)$.   Since the action of $\G$ on
$L^2(\A)$ preserves each subspace $L^2(\A_\gamma)$, the union of
the subspaces $L^2(\A_\gamma/\G_\gamma)$ must span $L^2(\A/\G)$.
So, what does $L^2(\A_\gamma/\G_\gamma)$ look like?  For this we
need a very precise picture of the action of  $\G$ on
$L^2(\A_\gamma)$.

Write $E$ for the set of edges of $\gamma$, and $V$ for the
set of vertices.  Then picking a trivialization of $P$
over the vertices of $\gamma$, we obtain an isomorphism
\ban      L^2(\A_\gamma) &\cong& L^2(G^E)    \\
&\cong& \bigotimes_{e\in E} L^2(G).  \ean
To see how $\G$ acts on the right-hand side, we use the
Peter-Weyl theorem.  Note that $G \times G$ acts on $G$ by
\[       (g_1,g_2)(h) = g_1 h g_2^{-1},  \]
giving a unitary representation of $G \times G$ on $L^2(G)$.
The Peter-Weyl theorem describes how $L^2(G)$ decomposes
into irreducible unitary representations of $G \times G$.  These
are all of the form $\rho_1 \tensor \rho_2$, where $\rho_1$ and
$\rho_2$ are irreducible unitary representations of $G$.  Let $R$
be a set containing one irreducible unitary representation of $G$
from each equivalence class.  Then the Peter-Weyl theorem says
that
\[     L^2(G) \cong \bigoplus_{\rho \in R} \rho \tensor \rho^\ast .\]
It follows that
\[     L^2(\A_\gamma) \cong \bigotimes_{e\in E} \bigoplus_{\rho \in R}
\rho \tensor \rho^\ast, \]
and in terms of this description, any $g \in \G$ acts on
$L^2(\A_\gamma)$ as the operator
\[     \bigotimes_{e\in E} \bigoplus_{\rho \in R}
\rho(g(s(e)) \tensor \rho^\ast(g(t(e))). \]
The reason is that the holonomy of $A$ along each edge
$e$ transforms in precisely the manner suited to applying
the Peter-Weyl theorem, as we saw in equation (\ref{tran}).

Next, using the associativity
of tensor product over direct sum, we obtain
\[     L^2(\A_\gamma) \cong \bigoplus_{\rho\in R^E}
\bigotimes_{e \in E} \rho_e \tensor \rho_e^\ast, \]
where $R^E$ is the set of all labellings of edges $e \in E$ by
representations $\rho_e \in R$.  Grouping the edges
by their source and target, this gives
\[     L^2(\A_\gamma) \cong \bigoplus_{\rho\in R^E}
\bigotimes_{v \in V} \left(\bigotimes_{e \in S(v)} \rho_e \tensor
\bigotimes_{e \in T(v)} \rho_e^\ast \right). \]
In these terms any
$g \in \G$ acts on $L^2(\A_\gamma)$ as the operator
\[ \bigoplus_{\rho\in R^E} \;
\bigotimes_{v \in V} \left( \bigotimes_{e \in S(v)}
\rho_e(g_v)  \otimes  \bigotimes_{e \in T(v)}
\rho_e^\ast(g_v) \right)   .\]
Thus we have
\[     L^2(\A_\gamma/\G_\gamma) \cong \bigoplus_{\rho\in R^E}
\bigotimes_{v \in V} {\rm Inv}\left(\bigotimes_{e \in S(v)} \rho_e \tensor
\bigotimes_{e \in T(v)} \rho_e^\ast \right), \]
where for any representation $\lambda$ of $G$, ${\rm
Inv}(\lambda)$ is the subspace of $G$-invariant vectors.
But ${\rm Inv}(\lambda_2 \tensor \lambda_1^\ast)$ is
isomorphic to the space
${\rm Hom}(\lambda_1, \lambda_2)$ of
intertwining operators from $\lambda_1$ to $\lambda_2$, so
finally we have
\[     L^2(\A_\gamma/\G_\gamma) \cong \bigoplus_{\rho\in R^E}
\bigotimes_{v \in V} {\rm Hom}\left(\bigotimes_{e \in T(v)}
\rho_e , \bigotimes_{e \in S(v)} \rho_e \right). \]

In other words, a complete set of gauge-invariant vectors in
$L^2(\A_\gamma)$ is obtained by labelling each edge of $\gamma$
with an irreducible unitary representation of $G$ and labelling
each vertex of $\gamma$ with an intertwining operator from the
tensor product of representations labelling incoming edges to
the tensor product of representations labelling outgoing edges.
If one unravels the logic of the proof, one sees that these are
special cases of the spin network states described more
explicitly above!

Our work has shown that spin network states span $L^2(\A/\G)$.
But in fact we have shown more.  For each embedded graph
$\gamma$ we obtain an orthonormal basis of
$L^2(\A_\gamma/\G_\gamma)$ by letting $\rho$ range over
all labellings of edges by irreducible unitary representations,
and, for each $\rho$, picking an orthonormal basis of
\[ {\rm Hom}\left(\bigotimes_{e \in T(v)}
\rho_e , \bigotimes_{e \in S(v)} \rho_e \right) \]
for each vertex $v$.  While it is a nuisance to assemble all
these orthonormal bases into a single orthonormal basis
for all of $L^2(\A/\G)$, for practical computations having
an orthonormal basis for each graph is usually sufficient.

Of course, for {\it really} practical computations one might need
a recipe for picking an orthonormal basis of the space of
intertwining operators
\[ {\rm Hom}\left(\bigotimes_{e \in T(v)}
\rho_e , \bigotimes_{e \in S(v)} \rho_e \right) .\]
For example, Br\"ugmann
\cite{Br2} has tried some computer simulations of quantum
gravity using spin networks, and computers are notorious for
wanting everything to be very explicit.  To pick such an
orthonormal basis, one needs to understand the representation
theory of $G$.  Luckily, the representation theory of $\SU(2)$ is
quite simple, so let us consider that case.   A more detailed
treatment can be found in the work of Rovelli and Smolin \cite{RS2}.

The irreducible unitary representations of $\SU(2)$ can be
labelled by their dimension $d = 1,2,3,\dots$, or equivalently
by their `spin' $j = 0,{1\over 2},1,\dots$, where $d = 2j+1$.
The tensor product of two such representations decomposes
as follows:
\[      j_1 \tensor j_2 = |j_1 - j_2|\; \oplus\; |j_1 - j_2|+1\; \oplus
\;\cdots \;\oplus\; j_1 + j_2 .\]
This means that for a {\it trivalent} graph $\gamma$ we
do not need to worry much about labelling the vertices with
intertwining operators.  Suppose, for example,
that a vertex has two incoming edges labelled with spins
$j_1$ and $j_2$, and one outgoing edge labelled $j_3$.
Unless the Clebsch-Gordon condition
\[      |j_1 - j_2| \le j_3 \le j_1 + j_2,\qquad
         j_1 + j_2 + j_3 \in \Z \]
holds, the representation $j_3$ will not appear as a summand of
of the tensor product $j_1 \tensor j_2$, and there is no way to
get a nonzero spin network state from this labelling.   If the
Clebsch-Gordon condition does hold, $j_3$ appears with
multiplicity 1 in $j_1 \tensor j_2$, so up to a constant factor
which one must fix in some standard way, there is no choice about
which intertwining operator to pick.

Graphs that are not trivalent can be reduced to the trivalent
case as follows.  Take each $n$-valent vertex and replace it in
ones mind by a $n$-leaved tree each of whose vertices is
trivalent. If the original vertex was $n$-valent, this tree will
have $n - 3$ new `internal edges'.   A basis of intertwining
operators for the original vertex is then given by all labellings
of these internal edges by spins satisfying the Clebsch-Gordon
condition.  Of course, there are many different trees with $n$
leaves, so there is a great deal of arbitrariness in this choice
of basis.  However, to change from one basis to another simply
requires repeated use of some matrices known as the $6j$ symbols,
which are familiar in $\SU(2)$ representation theory and whose
generalizations play an important role in topological quantum
field theory \cite{DJN,KL}.

Before returning to a more thorough treatment of the new
variables and the use of spin networks of quantum gravity, let us
say a little about the history of spin networks.   Because of the
immense difficulties they have had in quantizing gravity,
physicists have often considered the possibility that our picture
of spacetime as a manifold breaks down at the Planck scale.  This
has led to various attempts to reformulate physics in terms of
more discrete, or combinatorial, ideas.   In fact, spin networks
were invented in the early 1970s by Penrose \cite{Penrose} in one such attempt.
However, while our spin networks involve graphs {\it embedded} in a
pre-existing manifold that represents space, his spin networks
were {\it abstract} graphs labelled by representations and
intertwining operators.   In other words, rather than serving as
a tool for describing the geometry of a spacetime manifold, his
spin networks were intended as a purely combinatorial substitute
for a spacetime manifold.   The problem with this sort of radical
idea has always been bridging the gap between it and existing theories
of physics.   Thus the more recent discovery that spin networks
also arise naturally in attempts to quantize Einstein's equation
is quite intriguing.

If we succeeded in constructing quantum gravity as a field theory
on a manifold, we might simply decide to forget about more
combinatorial approaches to quantum gravity.  However, as we
shall see, the Hamiltonian constraint poses serious problems even
in the new variables formalism, so-called `ultraviolet problems',
which might be due to falsely extrapolating our picture of
spacetime as a manifold to arbitrarily small length scales.  Thus
it is still worth keeping in mind the possibility that some
combinatorial approach is the fundamental one.   For this reason,
any clues about the relation between `abstract' and `embedded'
spin networks are likely to be interesting.

Luckily, there are quite a few clues along these lines!   While
Penrose's ideas yield a purely combinatorial recipe for defining
an invariant of abstract $\SU(2)$ spin networks, this invariant
can also be described in terms of spin networks embedded in
$\R^3$.  One can obtain a generalized measure $\mu$ on the space
$\A$ of $\SU(2)$ connections on $\R^3$ by taking a finite Borel
measure $\mu_0$ on the subspace $\A_0$ of flat connections, and
defining
\begin{equation}
\int_{\A} \psi(A) d\mu = \int_{\A_0} \psi(A) d\mu_0 .
\label{flat} \end{equation}
Then if $\psi$ is a spin network state associated to some graph
$\gamma$, the integral $\int_\A \psi(A) d\mu$ does not depend on
the choice of embedding of $\gamma$,  essentially because the
holonomy of a {\it flat} connection along a path does not change
when the path is deformed.    Thus we obtain an invariant of
abstract spin networks.   If $\mu_0$ is a probability measure,
this is the same invariant that one can define
combinatorially following the ideas of Penrose (while carefully
adjusting his sign conventions \cite{RS2}).

A striking generalization of this idea arises from the work of
Witten \cite{Witten} on Chern-Simons theory and the Jones
polynomial, and its subsequent reinterpretation using
quantum groups by Reshtekhin and Turaev \cite{RT}.
Heuristically, Chern-Simons theory defines an
isotopy invariant of spin networks embedded in $S^3$
by means of a measure on the space of connections:
\begin{equation}  \int_{\A} \psi(A) e^{{ik\over 4\pi}\int_{S^3} \tr(A \we
dA + {2\over 3} A \we A \we A) } \D A   \label{cs} \end{equation}
Indeed, Witten was able by formal manipulations to compute these
invariants in certain cases.   Unfortunately, unlike equation
(\ref{flat}), this formula is difficult to make rigorous.  It
involves the purely formal `Lebesgue measure' $\D A$, and
replacing $\D A$ by the uniform generalized measure does not
help much, since the quantity integrated against it does not lie
in $\Fun(\A)$.  In fact, one cannot really compute these spin
network invariants using a generalized measure in our precise
sense of the term \cite{Baez3}.  One reason, though not the only
one, is that the invariants depend on a choice of extra structure
known as a `framing'.   Nonetheless, Reshetikhin and Turaev were able to
rigorously construct these isotopy invariants of framed embedded
spin networks by a combinatorial procedure that generalizes Penrose's
recipe to quantum groups.   In the case of links labelled by
the spin-${1\over 2}$ representation of $\SU(2)$, this invariant
is just the Kauffman bracket.

These ideas interact with the subject of quantum gravity in many
ways.  In Section 1 we cited some expository accounts of the
relation between Chern-Simons theory and quantum gravity in
4-dimensional spacetime.  However, spin networks and Chern-Simons
theory are also closely related to quantum gravity with
cosmological constant in 3-dimensional spacetime
\cite{Foxon,KL,PR,Roberts,R,TV,Witten2}.   In this case, quantum
gravity can be rigorously described {\it either} by starting with
Einstein's equation on a smooth 3-manifold, phrased in terms of
an appropriate version of the new variables, {\it or} purely
combinatorially in terms of a simplicial 3-manifold with edges
labelled by spins.   Better yet, although there is still a bit of work
left in proving this rigorously, it appears that the two
approaches are equivalent!   In short, in this toy model the
question as to whether spacetime is continuous or discrete has no
simple yes-or-no answer; in some sense, the answer is `both'.

It is natural to wonder if the same might be true in 4
dimensions.  There are a few hints here.  One the one hand,
Einstein's equation in the new variables formulation has a
simpler relative called $BF$ theory \cite{Baez3}, which is
much easier to canonically quantize.  On the other hand, Crane
and Yetter have combinatorially constructed a relatively simple
topological quantum field theory in 4 dimensions using spin
networks \cite{CKY}.  It is not known if these are equivalent,
though there are some clues suggesting it, such as the
relation of both to Chern-Simons theory.    This would be worth
understanding better, even though we expect full-fledged quantum
gravity in 4 dimensions to be more complicated.

\section{The New Variables}

Now let us return to general relativity and describe more
precisely how the new variables make it look more like other
gauge theories.  In its original formulation, general relativity
is all about a {\it metric} on spacetime, while gauge theories
are all about a {\it connection} on some bundle over spacetime.
Of course there is a connection involved in general relativity
too, the Levi-Civita connection, but this is traditionally
regarded as a subsidiary entity, since it can be computed
starting from the metric.  To emphasize the gauge-theoretic
aspects of general relativity,  one needs to rewrite it so a
connection has the starring role, and the metric appears as more
of a minor character.  There have been many different attempts to
do this, and for a more thorough exploration of them the reader
will have to turn elsewhere \cite{Peldan}.  Here we only describe
two: the Palatini formalism, and the Plebanski formalism.  The
latter is the one directly related to the `new variables', but
the former serves as a good warmup exercise.  This section makes
somewhat greater demands on the reader's acquaintance with
differential geometry, and also uses some variational calculus.
Luckily, there happens to be a textbook that explains most of
what we need \cite{BM}.  We will skip over all sorts of important
subtleties, which are discussed in Ashtekar's books \cite{Ash2,Ash3}.

In the Palatini formalism there will be two basic fields, a
connection and a `soldering form'.   The clever idea (in the
modern version of this approach) is to fix an oriented bundle
$\T$ over $M$ that is isomorphic to the tangent bundle $TM$, but
not canonically so.   We can think of $\T$ as a kind of
`imitation tangent bundle'.   Physicists usually call it, or any
of its fibers, the `internal space'.  We assume $\T$ is  equipped
with a Lorentzian metric $\eta$ --- the `internal metric'--- and
assume that the spacetime metric $g$ is obtained from $\eta$ via
an isomorphism $e \maps TM \to \T$.  In other words,
\begin{equation}     g(v,w) = \eta(e(v), e(w)) \label{g} \end{equation}
for any vector fields $v$ and $w$.
We may regard $e$ as a $\T$-valued 1-form, and then it is
called the soldering form.  We should add, in the interests of
cultural literacy, that $e$ is also called the `coframe field',
while $e^{-1}$ is known as the `frame field', `tetrad field',
or `vierbein'.

In Palatini formalism the two basic fields are this soldering
form $e$ and a connection $A$ on $\T$ preserving the metric
$\eta$, usually called a `Lorentz connection'.   One virtue of
this formalism is that it makes sense even when $e \maps TM \to
\T$ is {\it not} an isomorphism.  Thus the Palatini formalism
extends general relativity to certain cases where the metric $g$
is degenerate.

In computations it is handy to use  differential forms on $M$
taking values in the exterior algebra bundle $\Lambda \T$.  These
form an algebra, where the product (written as $\wedge$) is built
from the exterior product in $\Lambda \T$ together with the usual
wedge product of differential forms.  In complete analogy with
the volume form on an oriented Lorentzian manifold, the
orientation and internal metric on $\T$ give rise to an `internal
volume form', i.e.\ a section $\nu$ of $\Lambda^4\T$.   This in
turn yields a map from $\Lambda^4\T$-valued forms to ordinary
differential forms,  denoted by `$\tr$', and given by
\[    \tr(\nu \tensor \om) = \om \]
for any differential form $\om$ on $M$.
Now the curvature of the connection $A$ can, as usual, be
regarded as an $\End(\T)$-valued 2-form, but the internal metric
provides an isomorphism $\T \cong \T^\ast$, so we may think of
it as $\T \tensor \T$-valued, and then the fact that $A$ is
metric-preserving means the curvature actually takes values in $\Lambda^2
\T$.  We call this $\Lambda^2 \T$-valued 2-form $F$.

One can then obtain the vacuum Einstein equation from a variational
principle, starting with the `Palatini action' given by
\[
S_{Pal}(A,e) = \int_M \tr(e \we e \we F) .
\]
Let us sketch how this goes.  The idea is to compute the
variation $\delta S_{Pal}$, demand that it vanish for
all compactly supported variations $\delta A$ and $\delta e$,
and see what this implies.  We will need to use the wonderful
formula
\begin{equation}           \delta F = d_A \delta A,\label{var} \end{equation}
where $d_A$ denotes the exterior
covariant derivative of $\Lambda \T$-valued forms with respect
to the connection $A$, and $\delta A$ is treated as a $\Lambda^2
\T$-valued 1-form.  We begin by computing the variation, or
differential, of $S_{Pal}$ as a function of $A$ and $e$:
\ban   \delta S_{Pal} &=& \int \delta \tr(e \we e \we F) \\
&=& \int \tr (2\delta e \we e \we F +
 e \we e \we \delta F), \ean
using the standard rules for differentiation.  Using
equation (\ref{var}) we obtain
\[ \delta S_{Pal}
 = \int \tr (2 \delta e \we e \we F + e \we e \we d_A \delta A).
\]
Finally, integrating by parts as one always
does in these variational computations,
\[    \delta S_{Pal}
=  2\int \tr(\delta e \we e \we F - e \we d_A e \we \delta A).
\]
This only vanishes for {\it all} compactly supported $\delta A$,
$\delta e$ if
\[   e \we F = 0, \qquad e \we d_A e = 0.\]
These are our `classical equations of motion'.

When $e$ is an isomorphism, these equations are really just the
vacuum Einstein equation in disguise.  First, one can show that
in this case $e \we d_A e = 0$ implies  $d_A e = 0$.   Then,
using the isomorphism $e\maps TM \to \T$ to transfer $A$ to a
metric-preserving connection on $TM$, say $\Gamma$, the equation
$d_A e = 0$ implies that $\Gamma$ is torsion-free, hence equal to
the Levi-Civita connection of $g$.  This lets us translate the
equation $e \we F = 0$ into an equation about the curvature of
$\Gamma$, i.e., the Riemann tensor.  When we do so, we obtain the
vacuum Einstein equation!

The Plebanski formalism works in much the same way.  Indeed, at
first it  may seem like just an unnecessarily complicated version
of the Palatini formalism.  As noted by Ashtekar, however, the
{\it constraints} work out much more nicely when one tries to
canonically quantize the theory.   The Plebanski formalism is
also called the `self-dual' formalism,  because it takes
advantage of self-duality, a very special feature of 4
dimensions.  Recall that given a 4-dimensional real inner product
space $V$,  the Hodge star operator maps the second exterior
power of $V$ to itself:
\[          \star \maps \Lambda^2 V \to \Lambda^2 V.\]
Since $\star^2 = 1$, we can split any element $\om \in \Lambda^2 V$ into
a self-dual and an anti-self-dual part:
\[           \om = \om_+ + \om_-, \qquad  \star \om_{\pm} = \pm
\om_{\pm} .\]
Another way to think of this is as follows. The inner product
gives an isomorphism between  $\End(V) = V \tensor V^\ast$ and $V
\tensor V$, which restricts to an isomorphism between $\so(V)$
and $\Lambda^2 V$.   The splitting of $\Lambda^2 V$ into
self-dual and anti-self-dual parts then turns out to  correspond
to a splitting of the 6-dimensional Lie algebra $\so(V)$  into
two 3-dimensional ones.  Taking $V = \R^4$ this gives
\begin{equation}   \so(4) \cong \so(3) \oplus \so(3)  .\label{so}
\end{equation}
These simple facts have all sorts of ramifications for
4-dimensional physics and topology, especially in gauge theory,
where a connection on a Riemannian manifold having self-dual
curvature 2-form is automatically a solution of the Yang-Mills
equation, called an `instanton'  \cite{FU}.  When $M$ is compact,
the space of instantons modulo gauge transformations is
finite-dimensional, and one can extract powerful invariants of
the smooth structure of $M$ from this `moduli space' using
techniques developed by Donaldson and others \cite{DK}.  For
example, one can use these to show that $\R^4$ admits uncountably
many different smooth structures, unlike any other $\R^n$!
Seiberg and Witten have recently simplified some aspects of
Donaldson theory by further recourse to ideas from physics, but
self-duality still plays a key role.

The sort of self-duality relevant to the Plebanski formalism is
rather different, though with tantalizing relationships to the
above \cite{CS}.  First, given a Lorentzian rather than
Riemannian 4-manifold, one has $\star^2 = -1$ on 2-forms.  This
means that the eigenvalues of the Hodge star operator are $\pm
i$, so one needs to work with {\it complex-valued} 2-forms to
take advantage of self-duality.   When we tensor both sides of
equation (\ref{so}) with $\C$, we obtain
\[    \so(4,\C) \cong \ssl(2,\C) \oplus \ssl(2,\C)  . \]
For these reasons, the Plebanski formalism applies most naturally to
{\it complexified} general relativity, and some extra work is needed to
restrict to real-valued metrics.

Second, in the Plebanski formalism
self-duality first shows up not with respect to `honest' 2-forms, but
with respect to sections of $\Lambda^2 \C\T$, the second exterior
power of the complexified internal space.  The internal metric $\eta$
extends by complexification to a bilinear pairing on $\C\T$,
which we again call $\eta$, and this together with the
orientation give rise to an `internal' Hodge star operator $\ast$
acting on sections of $\Lambda^2 \C\T$.  We can thus split
sections of this bundle into self-dual and anti-self-dual parts:
\[           \om = \om_+ + \om_-, \qquad  \ast \om_{\pm} = \pm i
\om_{\pm} .\]
Now, the orthonormal frame bundle of $\C\T$ is a principal bundle
$P$ with structure group $\SO(4,\C)$.  Assume we have a spin
structure for $\C\T$, that is, a double cover $\tilde P$ of $P$
with structure group $\widetilde{SO}(4,\C) = \SL(2,\C) \oplus
\SL(2,\C)$.   Then $\tilde P$ is the sum $P_+ \oplus P_-$ of two
principal bundles with structure group $\SL(2,\C)$.   A
metric-preserving connection on $\C\T$ is usually called a
`complex Lorentz connection'.  By the above, a complex Lorentz
connection $A$ is equivalent to a pair of connections $A_{\pm}$
on $P_{\pm}$, which we call the self-dual and anti-self-dual
parts of $A$.   The curvature of $A_{\pm}$ is a 2-form $F_{\pm}$
taking values in the associated bundle ${\rm Ad} P_{\pm}$, but
sections of these bundles can be identified with self-dual
(resp.\ anti-self-dual) sections of $\Lambda^2 \C\T$.
The curvature of $A$ is a $\Lambda^2 \C\T$-valued 2-form
$F$ with $F = F_+ + F_-$.

The two basic fields in the Plebanski formalism are a
{\it complex-valued soldering form}, that is, 1-form on $M$ with values in
$\C\T$, and a {\it self-dual Lorentz connection} $A_+$, that is,
a connection on $P_+$.
The Plebanski action is given by:
\[   S_{Ple}(A_+,e) = \int_M \tr(e \we e \we F_+)
\]
This is very much like the Palatini action, and, much as before,
the classical equations of motion are
\[  e \we F_+ = 0,
\qquad e \we d_{A_+} e = 0.\]
These give Einstein's equation for complex-valued metrics on $M$
when $e$ defines an isomorphism between the complexified tangent
bundle of $M$ and $\C\T$.  To see this, first note that $e$ gives rise to a
complex-valued metric $g$ on $M$ by equation (\ref{g}).  Moreover,
we can use $e$ to transfer $A_+$ to a connection $\Gamma_+$ on the
complexified tangent bundle of $M$.   The equation $e \we d_{A_+}
e = 0$ then implies $\Gamma_+$ is torsion-free, hence equal to
the self-dual part of the Levi-Civita connection $\Gamma$
of $g$.   We can transfer $\Gamma$ back to $\C\T$ and get a
connection $A$ having $A_+$ as its self-dual part.  Then it turns
out that $e \we F_+ = e \we F$, so as in the Palatini formalism we
obtain Einstein's equation for $g$.

In terms of these `new variables', as they are often called, the
configuration space of general relativity is no longer the space
of all metrics on $S$.  Instead, it is the space $\A_+$ of all
connections on $P_+$ restricted to $S$, or more precisely, to
some fixed spacelike slice  $S_t \subset M$.  As in the metric
formulation, one can separate the equations of motion  into
evolutionary equations and constraints.   We shall not go through
the calculations here, but simply state some of the results
\cite{Ash2,Ash3}.   Recall that in the metric formulation of general
relativity, the analogs of the position and momentum are the
fields $q_{ij}$ and $p^{ij}$ on $S$.   In the new variables, the
analogs of position and momentum are instead the connection $A_+$
and the field $(e \we e)_+$ restricted to $S$.
Since $P_+$ is trivial when restricted to $S$, $A_+$ can
be identified with an $\ssl(2,\C)$-valued 1-form on $S$, usually
denoted simply by $A$.  Similarly,
$(e \we e)_+$ can be identified with an $\ssl(2,\C)$-valued
2-form on $S$, but physicists often use the isomorphism
\[           \Lambda^2 T^\ast S \cong TS \tensor \Lambda^3 T^\ast
S ,\]
given by the interior product, to treat it as a `densitized vector
field' on $S$ with values in $\ssl(2,\C)$, usually denoted by
$\tilde E$.   Actually, using coordinates,
they often write $A$ and
$\tilde E$ as $A_i^a$ and $\tilde E^{ia}$, where $a = 1,2,3$
indexes a basis of $\ssl(2,\C)$ that is orthonormal with respect to
the Killing form.

In these terms, the constraints in the new variables approach
are
\begin{equation}  \G^a = \partial_i \tilde E^{ia} + [A_i,\tilde E^i]^a ,
\label{con1} \end{equation}
\begin{equation}           \H_j = F^a_{ij} \tilde E^i_a , \label{con2}
\end{equation}
and
\begin{equation} \H = \epsilon_{abc} F_{ij}^c \tilde E^{ia} \tilde E^{jb},
\label{con3} \end{equation}
where $F$ is the curvature of $A$ and $\epsilon_{abc}$ is the completely
antisymmetric tensor with $\epsilon_{123} = 1$.

It is probably not obvious from our description where the
advantage of the self-dual formalism over the Palatini formalism
lies.  The key point is the construction of the field $\tilde E$
from $(e \we e)_+$.  One can similarly construct $\tilde E$ from
$e \we e$ in the Palatini formalism, but then $\tilde E$ winds up
being subject to extra constraints \cite{Ash3} which negate the
advantages of this approach.  In the self-dual
formalism, no conditions on $\tilde E$ need hold for it to come
from a complex soldering form $e$.

\section{Canonical Quantization}

Now let us try to canonically quantize gravity in terms of the new
variables, with an eye to the importance of spin networks.
The basic idea of DeWitt's approach goes over to this
context with only a few small modifications.  In its naive form,
the idea would be to define a Hilbert space $L^2(\A_+)$ and define
operators on this space by
\[     (\hat A_i^a \psi)(A) = A_i^a\psi(A), \quad
      (\hat {\tilde E}{}^i_a \psi)(A) = \hbar {\delta \psi \over \delta
A_i^a}(A).\]
(Again, we have suppressed the dependence of these operators on
the point in our spacelike slice.)   We would then substitute
these operators for $A$ and $\tilde E$ in formulas
(\ref{con1}-\ref{con3}), obtaining quantized constraints $\hat
\H$, $\hat \H_j$, and $\hat \G^a$.  Physical states should then
be annihilated by these constraints.

There are, however, a number of subtleties that we need to
address in order to do things right.   First there is the fact
that the Plebanski formalism most naturally describes complex
general relativity, and needs some adjustment to become a theory
of honest real-valued Lorentzian metrics.  This issue has been a
murky one for some time, and only now is it beginning to become
clear.  At the classical level one can simply impose extra
constraints saying that the metric is Lorentzian. This would not
be very nice in the quantum theory, though, because the whole
point of the new variables was to simplify the constraints!
Luckily it appears that at the quantum level we can deal with the
issue in quite a different way, namely by restricting our
attention to a space  $HL^2(\A_+)$ of holomorphic wavefunctions
on $\A_+$. Rather than really explaining this here, let us merely
note that it is closely related to the Bargmann-Segal formulation
of the quantum theory of a particle on a line \cite{Ash3,Ash4}.
The Bargmann-Segal formulation makes use of an isomorphism
between $L^2(\R)$ and the Hilbert space $HL^2(\C)$ consisting of
holomorphic functions that are square-integrable with respect to
a Gaussian measure.   A similar theory for complex Lie groups and
their compact real forms, due to Hall \cite{Hall},  gives an
isomorphism between a Hilbert space $HL^2(\SL(2,\C))$ and the
Hilbert space $L^2(\SU(2))$ defined using Haar measure.   A
generalization of this \cite{ALMMT} gives an isomorphism, the
`coherent state transform', between a Hilbert space $HL^2(\A_+)$
and the space $L^2(\A)$, where $\A$ is the space of connections
on an $\SU(2)$ bundle over $S$.  Thus the theory described in
Section 2 is relevant to quantum gravity even though $\SL(2,\C)$
is not compact.

Second, there is the issue of rigorously interpreting the
constraints and finding solutions to them.  When we try to
replace the fields $A$ and $\tilde E$ by $\hat A$ and $\hat
{\tilde E}$  in formulas (\ref{con1}-\ref{con3}), we run  into
difficulties.   One problem is that operators do not commute, so
different orderings of the same polynomial in the classical
fields can have different meanings at the quantum level.
The orderings given here seem best when one wants operators
on some space of functions on $\A_+$; elsewhere one may see
the opposite orderings \cite{Baez3}, but that is the natural
consequence of working dually on some space of generalized
measures.  A more profound problem is that $\hat A$ and $\hat
{\tilde E}$ are not really operator-valued functions, but only
operator-valued distributions --- one must integrate them against `test
functions' (really bundle sections) to obtain operators --- and
like ordinary distributions, the product of operator-valued
distributions is only defined under special conditions, or using
special tricks.

Luckily, the Gauss law constraint and diffeomorphism
constraint have simple geometrical interpretations which relieve
us of the need for making sense of the quantum analogs of
formulas (\ref{con1}) and (\ref{con2}).  At the classical level,
functions of the form $\int_S X_a \G^a$ generate all the flows on
the phase space $T^\ast \A$ that correspond to the action of
one-parameter groups of gauge transformations on $T^\ast \A$.  So
at the quantum level we can interpret the Gauss law constraint
as saying that a state is invariant under `small' gauge
transformations, that is, those lying in the component of $\G$
containing the identity.   One can show that any state $\psi \in
L^2(\A)$ that is invariant under small gauge transformations is
invariant under all gauge transformations.  As we have seen, there
is an ample supply of such states, and the space of these, which
we write as $L^2(\A/\G)$, is spanned by spin network states.

Similarly, any function on phase space of the form $\int_S N^i
\H_i$ generates a flow corresponding to the action of a
one-parameter group of diffeomorphisms.  Thus at the quantum
level we can interpret the diffeomorphism constraint
as saying that a state is invariant under the group
$\Diff_0(S)$ of small analytic diffeomorphisms, that is, those
lying in the  connected component containing the identity.   Here
we run into a problem.   It is easy to see that the only
element of $\Fun_0(\A)$ invariant under $\Diff_0(S)$ is the
function $1$; any other can be written as in equation
(\ref{form}) only for some nonempty graph $\gamma$, and then
there is a small diffeomorphism taking it to some other function,
in fact one orthogonal to it in $L^2(\A)$.  A simple
approximation argument then shows that no $\psi \in L^2(\A)$
except $\psi = 1$ is invariant under $\Diff_0(S)$!

In fact this is not as bad as it may seem.  Quite often in the
quantization of systems with constraints, the physical states are
not really vectors in the kinematical state space, but only
vectors in some larger topological vector space having the kinematical
state space as a dense subspace.   Consider a simple
example: the particle on the line, regarded as a particle
on the plane whose position $(q_1,q_2)$ is subject to the
constraint $q_2 = 0$.  Naively, one would start with the
kinematical state space $L^2(\R^2)$ and look for states $\psi$
satisfying the constraint $\hat q_2 \psi = 0$, i.e.,
\[              q_2\psi(q_1,q_2) = 0.\]
The only $L^2$ function on the plane satisfying this equation is
zero!  However, there are {\it distributions} on the plane
satisfying this equation.   The space of such solutions is
not itself a Hilbert space.  However, it has a subspace
isomorphic to $L^2(\R)$, given by the distributions of the form
$\psi(q_1)\delta(q_2)$.  This subspace is correct Hilbert space
for the particle on the line.

Similarly, while there are practically no solutions of the
diffeomorphism constraint living in $L^2(\A)$, there are plenty
of `distributional' ones.  In fact, there are plenty of
generalized measures on $\A$ invariant under small
diffeomorphisms of $S$.  The most interesting of these are the
gauge-invariant ones, since we would like to solve the Gauss law
constraint as well.   Note that the the inner product on
$L^2(\A/\G)$ sets up a chain of inclusions
\[        \Fun(\A/\G) \subset L^2(\A/\G)
\subset \Fun(\A/\G)^\ast ,\]
where $\Fun(\A/\G)$ is the gauge-invariant subspace of
$\Fun(\A)$, and its Banach space dual $\Fun(\A/\G)^\ast$
may be identified with the space of gauge-invariant
generalized measures on $\A$.   The group $\Diff_0(S)$ acts in a
consistent way on all these spaces.  A natural candidate for a
space of simultaneous solutions of the Gauss law and
diffeomorphism constraints is the space
\[      \Fun(\A/\G)^\ast_{inv} = \{ \mu \in \Fun(\A/\G)^\ast
\colon \; \forall g \in \Diff_0(S) \; g\mu = \mu \} .\]
This space is not itself a Hilbert space, but it may have some
subspace deserving to be called the `Hilbert space of
diffeomorphism-invariant states'.  On the other hand,
perhaps there are physically relevant diffeomorphism-invariant
states that are not contained in $\Fun(\A/\G)^\ast$, but only
in some still larger space containing $L^2(\A/\G)$.  An
example would be the `Chern-Simons state'.
To understand these would require further study of generalized
functions on $\A/\G$.

Anyway, at least a general recipe for finding elements of
$\Fun(\A/\G)^\ast_{inv}$ is known, as are many interesting examples
\cite{AL2,Baez}.  Among the most interesting are the
`knot states', which appeared already in a nonrigorous way
in the pioneering work of Rovelli and Smolin \cite{RS}. These are
most easily described using spin networks, although they were not
originally constructed that way \cite{Ash4}. Fix an isotopy class
$C$ of analytic knots and an  irreducible unitary representation
$\rho$ of $\SU(2)$.  This determines a set $S$ of spin network
states, namely the Wilson loops $\tr(\rho(T\exp \int_\alpha A))$
with $\alpha \in C$.   Since linear combinations of spin network
states are dense in $\Fun(\A/\G)$, any generalized measure $\mu
\in \Fun(\A/\G)^\ast$ is determined by its values on spin network
states.   We define the knot state $\mu$ by setting $\mu(\psi) =
1$ if $\psi \in S$ and $\mu(\psi) = 0$ if $\psi$ is a spin
network state not in $S$.  One must check, of course, that $\mu$
is a well-defined generalized measure, which involves proving  a
certain bound.  By construction $\mu$ is invariant under small
diffeomorphisms, so we have $\mu \in \Fun(\A/\G)^\ast_{inv}$.

Since a Wilson loop is just a special sort of spin network,  it
is natural to ask if this procedure generalizes to yield
`diffeomorphism-invariant spin network states'.  The idea would
be to use an arbitrary isotopy class of spin networks to obtain a
set $S$ of spin network states, and to define $\mu \in
\Fun(\A/\G)^\ast_{inv}$  by setting $\mu(\psi) = 1$ for $\psi \in S$ and
$\mu(\psi) = 0$ if $\psi$ is a spin network state not in $S$.
The problem is simply to prove that $\mu$ is a well-defined
generalized measure.  This seems to be true when we start with an
ambient isotopy class of {\it links} in $S$ labelled with
representations, but the general spin network case is more subtle
and still under investigation.

Now let us say a bit about the Hamiltonian constraint.
This is perhaps the most controversial aspect of the whole
subject, and certainly one of the most important ones:
if we find a way to rigorously treat the Hamiltonian constraint,
we will be quite close to a rigorous quantization of Einstein's
equation, but if not, the whole approach described here may
be fundamentally flawed, or at least in need of very new ideas.
A lot of work has been done on the Hamiltonian constraint, which
we cannot really do justice to here \cite{BGGP,Blencowe,Borissov,BP,G,JacSmo}.
Naively, the problem is to make sense of
\[ \hat \H(x) = \hbar^2 \epsilon_{abc} F_{ij}^c(x) {\partial \over \partial
A^{ia}(x)} {\partial\over \partial A^{jb}(x)}  \]
as a distribution on $S$ taking values in operators on
some space of holomorphic functions on $\A_+$.
Eventually, however, we want to find states
that are annihilated by this constraint together with the
Gauss law and diffeomorphism constraints.  Thus
we would be perfectly happy if we could first use the coherent
state transform to transfer the constraint to $L^2(\A)$,
and then find some subspace of $\Fun(\A/\G)^{\ast}_{inv}$
--- or some related space of solutions of the Gauss law
and diffeomorphism constraints --- that we could argue was
annihilated by the Hamiltonian constraint.   The problem is
that, in constrast to the Gauss law and diffeomorphism constraints,
there is no easy geometrical interpretation of the Hamiltonian
constraint in terms of connections on $S$ to fall back upon.

At first it might seem foolish to even {\it hope} for  a simple
geometrical 3-dimensional interpretation of the Hamiltonian
constraint.  After all, the Hamiltonian constraint expresses the
{\it 4-dimensional} diffeomorphism invariance of general
relativity; or in other words, it encodes the dynamics of the
theory.   It is all the more tantalizing, therefore, that there
are some hints of such an interpretation.  Rovelli and Smolin's
discovery of these  was one of the early successes of the loop
representation \cite{RS}, but making them precise has proved to
be very difficult.

Stripped of all nuance, the observation of Rovelli and Smolin
reduces to the following.  Given a knot $\alpha$ in $S$, the Wilson loop
\[   \psi(A) = \tr(\rho(T\exp \int_\alpha A)) \]
is a holomorphic function on $\A_+$.  Heuristically speaking,
when one applies the functional derivative $\partial/\partial
A_i^a(x)$ to $\psi$ one brings down a factor of the tangent vector
$\dot \alpha^i(t)$ if $\alpha(t) = x$.   So the double functional
derivative in $\hat \H(x)$ brings down a factor of
$\dot \alpha^i(t) \dot \alpha^j(t)$, which is symmetric in $i$
and $j$.   Since $F_{ij}^c$ is antisymmetric in $i$ and $j$,
one obtains
\[            \hat \H(x) \psi = 0 .\]

In fact, when one goes through the argument more carefully one
discovers that the double functional derivative of a Wilson
loop is very singular,
so the result $\hat \H(x) \psi = 0$ is a purely formal
one unless one can devise some regularization procedure to make
the argument rigorous.   This has not yet been achieved.   Still,
upon examination, the argument seems to suggest a 3-dimensional
geometrical interpretation of the Hamiltonian constraint as
some kind of `shift operator' generating the motion of a
Wilson loop, or more generally the edges of a spin network, along
its tangent vectors \cite{RS}.   For this reason, much effort
has been expended to understand the argument and find a context
in which it can be made rigorous. One would be very happy, for
example, to find by some heuristic argument a general formula for
the action of the Hamiltonian constraint on spin network states,
or perhaps `diffeomorphism-invariant spin network states', which
one could then subsequently justify by means of its good
properties.

In short, quantum gravity continues to provide mathematical
physics with challenging --- indeed quite frustrating ---
problems. However, let us conclude on a more upbeat note!
Even at the level of the kinematical state space $L^2(\A/\G)$,
there are some very intriguing applications of spin networks to
physics.  Classically, in the metric representation a state of
gravity is described by the metric on $S$ and its first time
derivative. We can rewrite interesting functions of the metric in
terms of the new variables, and then attempt to quantize them by
replacing $A$ and $\tilde E$ by their quantum versions
in these expressions, obtaining operators on $L^2(\A)$.   These
operators should commute with the action of $\G$, hence give rise
to operators on $L^2(\A/\G)$.   As the example of the Hamiltonian
constraint shows, carrying this out is by no means
straightforward in all cases.  However, Rovelli and Smolin
\cite{RS2} have considered the {\it area of a surface} and the
{\it volume of a region} in $S$, which are technically simpler,
and obtained explicit formulas for their quantum versions as
operators on $L^2(\A/\G)$, in terms of the spin network basis.
These operators turn out to have discrete spectrum: certain
multiples of Planck length squared for the area operator, and certain
multiples of the Planck length cubed for the volume operator.
In quantum theory, the spectrum of a self-adjoint operator
corresponds to the values the corresponding observable can
assume, so this is an indication that area and volume are
`quantized' in the very literal sense of assuming a discrete set of
values!  Moreover, there is a simple geometrical reason for
this fact.   To speak in rather oversimplified terms,
the area operator applied to a given spin network state
counts the number of points at which an {\it edge} of the spin
network intersects the surface in question, weighted by a factor
of $\sqrt{j(j+1)}\ell_p^2$, where $j$ is the spin labelling that edge.
Similarly, the volume operator applied to a spin network state counts
the number of {\it vertices} of the spin network contained in the region
in question, weighted by some function of their labellings by
intertwiners and the geometry of the incident edges.

Until we have a fully working theory of quantum gravity, and
understand how to take the other forces in account, it is
dangerous to take too seriously any physical predictions
a partial theory might make.  Moreover, this prediction of
discreteness of area and volume at the Planck scale is
absurdly hard to test with present technology!   Nonetheless,
the idea that the marriage of Einstein's equation and quantum
theory could make such a fascinating prediction should
serve as a kind of inspiration for mathematicians and physicists
working on quantum gravity.

\end{document}